\documentclass{article} 

\usepackage{graphicx} 
\usepackage{amsmath,url}
\usepackage{amssymb, mathtools, xcolor}

\newcommand{\R}{\mathbf{R}}
\newcommand{\B}{\mathbf{B}}

\newcommand{\x}{\mathbf{x}}
\newcommand{\y}{\mathbf{y}}

\newcommand{\M}{\mathbf{M}}

\newcommand{\J}{\mathcal{J}}

\def\HH{\mathbf{H}}
\def\nens{\textsc{n}_{\rm ens}}

\usepackage{todonotes}

\usepackage[numbers]{natbib}
\usepackage{hyperref}

\begin{document}

\title{Weakly-constrained 4D Var for downscaling with uncertainty using data-driven surrogate models}

\maketitle
\author{Philip Dinenis $^1$ (pdinenis@anl.gov),}
\author{Vishwas Rao$^1$,}
\author{Mihai Anitescu$^1$}\\

1. Argonne National Laboratory, Lemont, IL

\date{March 3, 2025}

\begin{abstract}{Dynamic downscaling typically involves using numerical weather prediction (NWP) solvers to refine coarse data to higher spatial resolutions. Data-driven models such as FourCastNet have emerged as a promising alternative to the traditional NWP models for forecasting. Once these models are trained, they are capable of delivering forecasts in a few seconds, thousands of times faster compared to classical NWP models. However, as the lead times, and, therefore, their forecast window, increase,  these models show instability in that they tend to diverge from reality. In this paper, we propose to use data assimilation approaches to stabilize them when used for downscaling tasks.  Data assimilation uses information from three different sources, namely an imperfect computational model based on partial differential equations (PDE), from noisy observations, and from an uncertainty-reflecting prior. In this work, when carrying out dynamic downscaling, we replace the computationally expensive PDE-based  NWP models with FourCastNet in a ``weak-constrained 4DVar framework" that accounts for the implied model errors. We demonstrate the efficacy of this approach for a hurricane-tracking problem; moreover, the 4DVar framework naturally allows the expression and quantification of uncertainty. We demonstrate, using ERA5 data, that our approach performs better than the ensemble Kalman filter (EnKF) and the unstabilized FourCastNet model, both in terms of forecast accuracy and forecast uncertainty.
}
\end{abstract}

\newpage

\section{Introduction}
Downscaling refers to methods used in atmospheric science to refine coarse-scale climate or weather data and estimate the state at finer temporal or spatial resolutions \cite{hewitson1996climate}. This paper will consider spatial downscaling only, which can be understood as a super-resolution problem. This problem often arises in climate projections where Global Climate Models (GCMs) output is too coarse for many applications. These GCM outputs can be refined using a nested high-resolution regional climate model (RCM), which typically takes the form of high-resolution numerical weather prediction (NWP) -based solvers. However, recent advances in machine learning methods have given rise to data-driven forecasting methods such as FourCastNet \cite{pathak2022fourcastnet}, 
that are orders of magnitude faster than NWP in short to medium term \cite{pathak2022fourcastnet, bi2023panguweather, weyn2019can}. 
 These data-driven methods offer similar accuracy for forecasting on the order of a few days, after which they lose coherence with the ground truth. 
 In this work, we use data assimilation (DA) to stabilize data-driven models when carrying out dynamic downscaling tasks while using model error to account for the errors from the data-driven forecasting methods. Furthermore, using DA for downscaling allows us to systematically quantify the uncertainties. 

In the context of downscaling, two main approaches have been developed to produce an estimate of the fine-scale information: statistical downscaling \cite{wilby2013statistical} and dynamical downscaling \cite{tapiador2020regional}. In statistical downscaling, high-resolution reconstructions are obtained by estimating the statistical relationship between the fine and coarse information. This can be done using techniques such as weather classification schemes, regression models, and weather generators \cite{wilby2004guidelines}. These techniques are computationally inexpensive, however they do not use a model for the dynamics and thus can oversimplify the complex interactions between the coarse and fine scales \cite{zorita1999analog}. Dynamical downscaling involves a nested high-resolution regional climate model (RCM). These are more computationally expensive and thus limited to certain regions. The RCM solves the forward problem using the boundary conditions of the GCM. In the context of future climate projections, the initial condition that the RCM may also be the output of the GCM. When initialized on these coarse states, RCM outputs can produce unreliable results at first. They are typically run for a long period, called the 'spin-up' or 'trash period,' only after which the outputs are used \cite{https://doi.org/10.1029/2019MS001945}. This can be expensive and requires stability in the RCM. Additionally, the computational expense of RCMs can limit the feasibility of ensemble methods for uncertainty quantification. They also require careful boundary handling to avoid inconsistent physics, spurious waves, or other artifacts. Data assimilation techniques can also be used 
in traditional dynamical downscaling to improve the estimate of the initial condition of the RCM. Then, the fine-scale trajectory is found by solving the forward problem with improved initialization \cite{peng2010application, zupanski2011prototype}. 

Machine learning (ML) techniques are emerging as a popular tool for weather and climate assessment tasks and, in the context of this paper, for statistical downscaling tasks. Downscaling can be viewed as a supervised learning task, where the goal is to learn the variables on the fine grid (the target), from the variables on the coarse grid (the input). Most machine learning studies aim to directly learn the mapping between the coarse grid input and the fine grid target. A wide variety of ML approaches such as  Convolutional Neural Networks (CNNs), U-Nets, convolutional-autoencoders, and Fourier neural operators \cite{wang2019end, adewoyin2021tru, babaousmail2021novel, yang2023fourier} have been used to downscale climate data. While these approaches show tremendous promise particularly in regards to the compute time,
 a key element missing in these approaches is the ability to quantify the uncertainty in the downscaling \cite{watt2024generative}
 There are a few notable exceptions to this issue, such as the approaches based on normalizing flows \cite{groenke2020climalign, winkler2024climate}, which provide a systematic framework for uncertainty quantification in the downscaled data. Moreover, generative models tend not to incorporate the underlying dynamics and focus on static matching.

In this work, we aim to provide a method to carry out dynamic downscaling with ML surrogates, such as FourCastNet -- to harness their significant speed-up relative to NWP approaches while also addressing their potential instability: their divergence from the actual trajectory, which makes the spin-up philosophy of RCM not applicable to their case.  To this end, we solve an inverse problem for the entire fine-scale trajectory of the ML surrogate while enforcing consistency with the coarse-scale information on the entire forecast horizon. This is done by formulating downscaling as a data assimilation problem, where the data assimilated includes the coarse scale information. This approach incorporates aspects of both dynamical and statistical downscaling. It gives us a systematic means of quantifying uncertainty as we estimate a posterior on the fine-scale information conditioned on the coarse scale. In this sense, our method resembles statistical downscaling but has the advantage that a dynamic model constrains it. Additionally, since FourCastNet can be efficiently run on a global domain, we do not need to treat the coarse scale as a boundary condition. 



In its classical use, DA uses information from three different sources: an imperfect computational model based on differential equations in the case of NWP, noisy observations (sparse snapshots of reality), and an uncertain prior (which encapsulates our current knowledge of reality). DA integrates these three sources of information and the associated uncertainties in a Bayesian framework to provide the posterior, i.e., the probability distribution of the trajectory conditional on the uncertainties in the model and observations \cite{Daley_B1993,Kalnay_B2003,LeDimet_1986}. This technique is used extensively in numerical weather prediction (NWP) to make accurate predictions about the state of the atmosphere and oceans \cite{Kalnay_B2003, Navon_1998}. Please see \S\ref{sec:da} for more details about DA. 

The emergence of data-driven weather forecasting, particularly the use of high-quality surrogates, is a promising alternative to traditional numerical weather prediction (NWP) methods in recent years. 
Once such models are trained, they can potentially deliver forecasts at a much lower computational cost. FourCastNet, introduced by Pathak et al., \cite{pathak2022fourcastnet} is an example of such an approach we will use in this work.
 FourCastNet is a global data-driven weather forecasting model that uses a Fourier transform-based token-mixing scheme with a vision transformer backbone. It provides accurate short to medium-range global predictions at $0.25^{\circ}$ resolution and can generate a week-long forecast in less than two seconds\cite{pathak2022fourcastnet}. Pangu-Weather, developed by Bi et al., employs 3D deep neural networks to capture atmospheric dynamics. It uses a hierarchical temporal aggregation method to reduce cumulative forecast errors and has shown competitive performance against traditional NWP models\cite{bi2023panguweather}. GraphCast, presented by Lam et al., utilizes graph neural networks for weather forecasting. It has demonstrated strong performance across various atmospheric and surface variables over multiple lead times\cite{lam2023graphcast}. Many recent works use ML-based surrogates to forecast weather\cite{keisler2022forecasting, price2023gencast, chen2023fengwu, bodnar2024aurora}. While ML-based surrogates such as FourCastNet, GraphCastNet, and Pangu weather provide rapid forecasts, 
 we note that they are trained on reanalysis data and must be re-trained when their forecasts diverge from reality. This work integrates the pretrained data-driven models, such as FourCastNet, with observational data in a variational DA framework to correct the initial conditions and provide rapid forecasts. In turn, it greatly mitigates the computational costs associated with dynamic downscaling 

The primary contributions of this work are as follows: 
\begin{itemize}
\item We use DA for downscaling by leveraging a pretrained, fast, and differentiable model (FourCastNet) in a weak-constraint 4DVar framework to obtain the fine-resolution downscaled over a given downscaling window.
\item We quantify the uncertainty in these downscaled data by using the inverse of the Hessian to approximate the covariance at the maximum-aposteriori point (MAP) (which is also the Laplace approximation).
\item We present extensive numerical results with the ERA5 dataset for downscaling a hurricane and compare the results with that of EnKF and an unassimilated FourCastNet prediction.
\end{itemize}
The rest of the work is organized as follows: in \S \ref{sec:fcn}, we describe the FourCastNet model; in \S \ref{sec:da}, we review the principal data assimilation approaches; in \S\ref{sec:method} we describe our methodology; and in \S\ref{sec:experiments} we demonstrate the efficacy of our methods in numerical experiments using Hurricane Michael as an illustrative example.

\section{FourCastNet}\label{sec:fcn}
In this work, we will use FourCastNet \cite{pathak2022fourcastnet} as a surrogate of the fine-scale dynamics. We proceed to describe its features. 
FourCastNet \cite{pathak2022fourcastnet}  is a deep-learning-based global weather forecasting model, able to generate accurate short to medium-range global forecasts at a $0.25^{\circ}\times0.25^{\circ}$ spatial resolution (approximately 28km$\times$28km at the equator), 
 at a six-hour temporal resolution. FourCastNet considers twenty atmospheric variables, including water vapor, temperature, and wind velocity at different vertical levels. Together, these provide a detailed description of the atmospheric state, critical for predicting severe weather events and supporting applications like wind energy resource planning. A full description of the FourCastNet input and output variables is found in Table \ref{tab:vars}. Each variable uses normalized units, such that they have mean 0 and standard deviation 1 on the historical training set. We will use these units unless other units are specified.

\begin{table}[ht]
    \centering
    \begin{tabular}{|c|c|}
        \hline
        \textbf{Variables} & \textbf{Vertical Levels} \\ \hline
        T & 500hPa, 850hPa   \\ \hline
        U, V & 500hPa, 850hPa, 1000hPa \\ \hline
        Z & 50hPa, 500hPa, 850hPa, 1000hPa  \\ \hline
        RH & 500hPa, 850hPa \\ \hline
        sp, mslp, $U_{10}$, $V_{10}$, $T_{2m}$ & surface  \\ \hline
        TCWV & integrated  \\ \hline
    \end{tabular}
    \caption{Table of input and output variables at each grid point. The variables with their corresponding abbreviations are: temperature (T), zonal wind velocity (U), meridional wind velocity (V), geopotential (Z), relative humidity (RH), surface pressure (sp), mean sea level pressure (mslp), zonal and meridional wind velocities 10m from surface ($U_{10}$ and $V_{10}$ respectively) , temperature 2m from surface ($T_{2m}$) and total column water vapor (TCWV). T, U, V, Z and RH are modeled at a various pressure levels, which represent vertical layers in the atmosphere. Note that FourCastNet is a time-independent model and does not take time or time of day as an input.}
\label{tab:vars}
\end{table}

One of FourCastNet's most notable advantages is its exceptional computational speed. It can generate a week-long forecast in under two seconds, roughly 45,000 times faster than traditional NWP models such as the ECMWF Integrated Forecasting System (IFS) \cite{pathak2022fourcastnet}. The computational efficiency of FourCastNet enables the rapid generation of extensive ensemble forecasts with thousands of ensemble members, enhancing probabilistic forecasting and uncertainty quantification. Moreover, the artificial neural network structure of FourCastNet allows for automatic differentiation with respect to its inputs. This enables the use of FourCastNet as a surrogate in variational inverse problems.

FourCastNet maps from a parameterized atmospheric state to a predicted state six hours later. We denote this map by $\mathcal M$, 
$$\mathcal M : \mathbb R^{d_x} \rightarrow \mathbb R^{d_x}$$
$$d_x := n_{\rm var}\times n_{\rm lat}\times n_{\rm lon} =20\times 720 \times 1440 = 20,736,000$$
In this way, it can be understood as a solution operator that acts as a surrogate for atmospheric physics. This operator is learned purely from historical reanalysis data. The training set is a subsample of the years 1979 to 2015 inclusive, taken from the ECMWF Reanalysis v5 (ERA5) dataset, produced by Copernicus Climate Change Service (C3S) at ECMWF \cite{ERA5}. FourCastNet is based on the Adaptive Fourier Neural Operator architecture, which is a kind of a vision transformer with Fourier-based token mixing \cite{pathak2022fourcastnet,guibas2022adaptivefourierneuraloperators}. Other neural operator architectures have also been successfully used to learn surrogate solution operators to partial differential equations. \cite{li2021fourierneuraloperatorparametric}

\section{Data Assimilation}\label{sec:da}
While this paper focuses on downscaling and stabilizing data-driven surrogate models to this end, the primary tool used will be DA. We proceed to describe it in a way that expresses how the elements of our approach are chosen and integrated. 

DA is the process of fusing information from priors, imperfect model predictions, and noisy data to obtain a consistent description of the actual state $\x^{\rm true}$ of a physical system \cite{Daley_B1993, Kalnay_B2003, Sandu_2011, Sandu_2005}. The best estimate is the analysis $\x^{\rm a}$. The prior information encapsulates our current system knowledge; it typically consists of a background estimate of the state $\x^{\rm b}$ and the corresponding background error covariance matrix $\mathbf{B}$. 
Knowledge about the physical laws that govern the evolution of the system is incorporated into a map 
$\mathcal{M}_{\tau_a \rightarrow \tau_b}(\cdot)$ which in the case of NWP is built from PDE approximations. 
The resulting model evolves an initial state $\x_0 \in \mathbb{R}^n$ at the initial time $t_{ 0 }$ to states $\x_{i} \in \mathbb{R}^n$ at future times $t_{i}$ through the map $\mathcal{M}$: 
\begin{equation}
\label{eqn:genmodel}
 \x_{i} = \mathcal{M}_{t_0 \rightarrow t_{i}} \left(\x_0\right), \quad  i=1,\cdots, N.
\end{equation}
Observations are noisy snapshots of reality available at discrete time instances. Specifically, measurements $\y_{i} \in \mathbb{R}^m$ of the physical state $\x^{\rm true}\left(t_{i}\right)$ are taken at times $t_{i}$,
\begin{equation}
\label{eqn:genobservation}
\y_{i} = \mathcal{H}(\x_i) + \varepsilon_i, \quad  \varepsilon_i \sim \mathcal{N}(\mathbf{0},\R_i),
\quad i=1,\cdots, N,
\end{equation}
where the observation operator $\mathcal{H} : \mathbb{R}^n \to \mathbb{R}^m$ maps the model state space onto the observation space. The random observation errors $\varepsilon_i$ are assumed to have normal distributions here. In general, both the model and the observation operators are nonlinear. 
All models are imperfect; for example, NWP models are laden with uncertainties due to incomplete knowledge of the physics associated with sub-grid modeling, errors in boundary conditions, accumulation of numerical errors due to temporal discretizations, inexact forcings, etc. Recent applications have called for aggregating these uncertainties into a model error component \cite{Glimm_2004,Orrell_2001,Palmer_2005}.

Two approaches to data assimilation have gained widespread popularity: ensemble-based estimation \cite{evensen2003ensemble} and variational methods \cite{Derber_1989,Sasaki_B1970}. The ensemble-based techniques are rooted in statistical theory, whereas the variational approach is derived from optimal control theory \cite{evensen2022data}. 
 The variational approach formulates data assimilation as a nonlinear optimization problem constrained by a numerical model \cite{rabier2005overview, le1986variational, navon2009data}. The initial conditions (as well as boundary conditions, forcing, or model parameters) are adjusted to minimize the discrepancy between the model trajectory and a set of time-distributed observations. In real-time operational settings, the data assimilation process is performed in cycles: observations within an assimilation window are used to obtain an optimal trajectory, which provides the initial condition for the next time window, and the process is repeated in the subsequent cycles \cite{law2015data, asch2016data}. 


DA in the variational approach (4DVar) is often implemented as a strong constraint algorithm (SC4DVar). The assumption that the model is perfect or that the model error is small enough relative to other errors in the system to be ignored. However, the error in the model is often large enough to be non-negligible, affecting the SC4DVar scheme and resulting in an analysis inconsistent with the observations. The effect of model error is even more pronounced when the assimilation window is large \cite{laloyaux2020improving, ren2021effects}.
 Weak constraint 4DVar (WC4DVar) \cite{Tremolet_2006, Tremolet_2007} relaxes the `perfect model' assumption and assumes that model error is present in each time step. However, both SC4DVar and WC4DVar are computationally expensive, and using them in operational settings can be challenging. An alternative is to use a linearized version called the incremental 4DVar to reduce the computational burden \cite{bannister2017review}. Both SC4DVar and WC4DVar require repeated solutions to the forward and adjoint models -- both of which are based on large-scale partial differential equations (PDE) when used within NWP \cite{Akella_2009, Cardinali_2014, Hansen_2002}. Below, we describe the SC4DVar and WC4DVar since both these approaches will be relevant to the downscaling studies performed in this work.






\subsection{Strong-Constraint 4D-Var}
 SC4DVar processes simultaneously all observations at all times $t_{1},t_{2}, \dots, t_{N}$ within the assimilation window. The control parameters are typically the initial conditions $\x_0$, which uniquely determine the system's state at all future times, assuming that the model \eqref{eqn:genmodel} perfectly represents reality. The background state is the prior best estimate of the initial conditions $\x_0^{\rm b}$ and has an associated initial background error covariance matrix $\mathbf{B}_0$. The 4DVar problem provides the estimate $\x_0^{\rm a}$ of the true initial conditions as the solution of the following optimization problem
\begin{subequations}
\label{eqn:L2-4dvar}
\begin{eqnarray}
\label{eqn:ip}
&&~ \x_0^{\rm a}  =  \underset{\x_0} {\text{ arg\, min}}~~ \J\left(\x_0\right) \qquad
 \text{subject to}~ \text{\eqref{eqn:genmodel}},\\
\label{eqn:fdvar-cf}
&&~ \mathcal{J}\left(\x_0\right) := \frac{1}{2}\| \x_0 - \x_0^{\rm b} \|_{\B_0^{-1}}^2 +
\frac{1}{2} \, \sum_{i=1}^{N}\| \mathcal{H}(\x_i) - \y_i \|_{\R_i^{-1}}^2 \,.
\end{eqnarray}
\end{subequations}
The first term of \eqref{eqn:fdvar-cf} quantifies the departure of the solution $\x_0$ from the background state $\x_0^{\rm b}$ at the initial time $t_0$. The second term measures the mismatch between the forecast trajectory (model solutions $\x_{i}$) and observations $\y_{i}$ at all times $t_{i}$ in the assimilation window. The covariance matrices $\mathbf{B}_0$ and $\mathbf{R}_{i}$ are predefined, and their quality influences the accuracy of the resulting analysis.
%

 The minimizer of \eqref{eqn:ip} is computed iteratively using gradient-based numerical optimization methods. First-order adjoint models provide the gradient of the cost function \cite{Cacuci_B2005}, while second-order adjoint models provide the Hessian-vector product (e.g., for Newton-type methods) \cite{Sandu_2008}. The methodology for building and using various adjoint models for optimization, sensitivity analysis, and uncertainty quantification is discussed in \cite{Sandu_2005, Cacuci_B2005, Cioaca_2012, Rao_2015}.

 \subsection{Weak constrained 4DVar} \label{ss:4DVar}
We describe the WC4DVar formulation in this subsection. The work in \cite{Tremolet_2006} describes three different formulations of WC4DVar, the control variables being distinguishing between the formulations. Since the model representing the atmosphere is allowed to be imperfect, 
we allow for model error as follows:
\begin{align}\label{eqn:modelerror}
 \mathbf{x}_i = \mathcal{M}_{t_{i-1} \rightarrow t_i} \left(\mathbf{x}_{i-1} \right) + \boldsymbol{\eta}_i\, , \;   \boldsymbol{\eta}_i \sim \mathcal{N}(\mathbf{0},\mathbf{Q}_i).
\end{align}
Here, $\mathcal{M}_{t_{i-1} \rightarrow t_i}$ represents the model describing the evolution of the atmospheric flow from time $t_{i-1}$ to $t_i$ and $\boldsymbol{\eta}_i$ is the model error in the time-step; assumed here to be Gaussian.
%
This formulation is called ``weakly constrained 4DVar'' since the estimated trajectory $\mathbf{x}$ will not precisely satisfy the recursion defined by $\mathcal{M}$. For reasons related to the size of the problem stated in \cite{Tremolet_2006}, one sometimes makes simplifying assumptions in the representation of the control variable. For example, instead of solving for the entire state trajectory, we may solve for the states at the observation time points alone. This alleviates the computational challenge considerably.
In our case, we choose the state variable $\x$ as the control variable. The WC4DVar cost function is then given by:
\begin{align}
\label{eqn:wc4dvar-cost}
 \mathcal{J}(\mathbf{x}_0, \cdots,\mathbf{x}_n ) &= \frac{1}{2} \|\mathbf{x}_0 - \mathbf{x}_b\|^2_{\mathbf{B}^{-1}_0} + \frac{1}{2}\sum_{i=0}^{n}\,\|\mathcal{H}_i(\mathbf{x}_i) - \mathbf{y}_i\|^2_{\mathbf{R}^{-1}_i} + \nonumber \\
& \frac{1}{2} \sum_{i=1}^n \,\|\mathbf{x}_i - \mathcal{M}_{t_{i-1} \rightarrow t_i}(\mathbf{x}_{i-1})\|^2_{\mathbf{Q}^{-1}_i} \,.
\end{align}
As an aside, we note that the error distributions need not be Gaussian, provided we can write (the negative log-likelihood, \eqref {eqn:posterior_prop}) 
 $\mathcal{J}$ explicitly; the approach would largely stay the same. The WC4DVar optimization problem can then be written as 
\begin{equation}\label{eqn:ip_weak}
 \x^{\rm a}  =  \underset{\x} {\text{ arg\, min}}~~ \J\left(\x\right) \qquad,
\end{equation}
where $\x\coloneqq (\mathbf{x}_0, \cdots,\mathbf{x}_n ) $\,.
The optimization problem in \eqref{eqn:ip_weak} can be solved in memory-constrained environments using the LBFGS method, which requires the evaluation of gradients \cite[Section 7.2]{nocedal1999numerical}. The gradient can be written as
\begin{subequations}
\label{eqn:AugLagGrad}
\begin{eqnarray}
\label{eqn:AugLagGrad0}
\nabla_{\x_0} \mathcal{J}&=&\B_0^{-1} \left(\x_0 - \x_0^{\rm b}\right)  -  \M^{\rm T}_{0,1} \, \mathbf{Q}_0^{-1} \left(\x_1 - \mathcal{M}_{0,1}\left(\x_0\right) \right)\,, \\ [5pt]
\label{eqn:AugLagGrad1} 
        \nabla_{\x_{\rm k}}\mathcal{J} &=&\mathbf{H}_{\rm k}^{\rm T} \R_{\rm k}^{-1} \left(\mathcal{H}\left(\x_{\rm k}\right) -\y_{\rm k} \right) \\
&& +  \mathbf{Q}^{-1}_{{\rm k}-1}\left(\x_{\rm k} - \mathcal{M}_{\rm k-1,k}\left(\x_{\rm k-1}\right)\right)  \nonumber  \\ 
 && -  \mathbf{M}^{\rm T}_{\rm k,\rm k+1}\, \mathbf{Q}^{-1}_{\rm k} \left(\x_{\rm k+1} - \mathcal{M}_{\rm k, k+1} \left(\x_{\rm k}\right)\right)\,, \nonumber  
 \nonumber
  \quad \ {\rm k} = 1, \dots,  N-1, \\ [5pt]
\label{eqn:AugLagGrad2} 
\nabla_{\x_{\rm N}}\mathcal{J} &=& \mathbf{H}_{\rm N}^{\rm T} \R_{\rm N}^{-1}\left(\mathcal{H}\left(\x_{\rm N}\right) -\y_{\rm N} \right) +  \mathbf{Q}_{\rm N -1}^{-1} \left( \x_{\rm N} - \mathcal{M}_{\rm{N}-1, \rm N}\left(\x_{\rm N-1}\right) \right)\,.
\end{eqnarray}
\end{subequations}
Where, $\mathbf{M}_{\rm k,\rm k+1}$ represents the tangent linear opeator and $\mathbf{M}^{\rm T}_{\rm k,\rm k+1}$ represents the adjoint operator. We assume
that the observation operator $\mathcal{H},$ model error covariance $\mathbf{Q},$ measurement error variance $\mathbf{R},$ and model $\mathcal{M}$ are independent of time. This approach matches the surrogate we use, as  FourCastNet operates under this time-independent assumption. It does not take time or time of day as an input and is trained as a 'stationary model', just minimizing the loss in single-step forecasts (with additional fine tuning on two-step forecasts).  

\section{Methodology}\label{sec:method}
This section details how we incorporate FourCastNet into the weak constraint 4DVar framework and perform uncertainty quantification on the downscaled results. We also describe our methodology for incorporating FourCastNet into an Ensemble Kalman filter (EnKF) for latent-state estimation; this is used as a comparison for WC4DVAR, our primary method.

\subsection{Model error of FourCastNet}
\label{sec:Q_est}
To use FourCastNet 
in the weakly constrained framework of \S \ref{ss:4DVar} we need to estimate its model error 
 to provide the parameters $\mathbf{Q}_i$ in the WC4DVAR cost function (see equation \ref{eqn:wc4dvar-cost}) and in the stochastic forecast step in the EnKF (see \ref{eqn:forecast-step}). In the formulations of WC4DVAR and EnKF, the model error is assumed to be Gaussian with mean zero \cite{Tremolet_2006, evensen2022data}. Using historical data, we can compute the sample covariance of the model error in FourCastNet predictions. 

We assume the model error covariance has no time dependence, i.e., that $\mathbf Q_0,..., \mathbf Q_N $ are all equal and denote this as $\mathbf Q$. To simplify the estimation and the subsequent computations, we assume $\mathbf Q$ is diagonal so that there is no covariance between the error at different points nor between different atmospheric variables. We additionally assume that the model error variance depends only on the predicted atmospheric variable and is independent of the location of the grid point. Thus $\mathbf{Q}$ is diagonal with entries taking values only in the set $\{q_1,...,q_{20}\}$ representing the model error variances of the 20 atmospheric variables. We calculate the sample variance of atmospheric variable $v$ using historical data from random samples of times of ERA5 data that was not used in the training of FourCastNet nor the testing in the numerical experiments section of this paper.
$$q_v = \frac{1}{n_{\rm lat}n_{\rm lon}|S|-1}\sum_{t \in S}\sum_{p=1}^{n_{\rm lat}n_{\rm lon}}(\mathcal M(\x^{\rm true}_{\rm t})_{p,v}- (\x^{\rm true}_{\rm{t+6h}})_{p,v})^2,$$
where indices $p$ and $v$ refer to grid point and atmospheric variable respectively, $S$ is a randomly drawn subset of out-of-sample times and $\x_t^{true}$ represents the ERA5 data at a time $t$. $q_v$ is the model error variance for atmospheric variable $v$.

We chose a diagonal model error covariance matrix for ease of computation, invertibility, and reduction of spurious correlations. We note that a less sparse choice of $\mathbf Q$ can be more informative by better capturing the spatial correlations; however, we defer that to future extensions of this work.


\subsection{Optimization of WC4DVAR}
\label{ss:optimization}
To find the minimizer of the WC4DVAR cost function $ \mathcal{J}$ (equation \ref{eqn:wc4dvar-cost}, we use the limited-memory Broyden–Fletcher–Goldfarb–Shanno algorithm (L-BFGS) \cite[Section 7.2]{nocedal1999numerical}. L-BFGS, a quasi-Newton method, approximates the inverse Hessian using a history of update vectors. This method circumvents storing the full Hessian matrix and is well-suited for our high-dimensional optimization problem \cite{Liu1989}. 

We denote the value of the optimization variable at the $i$th iteration as $\x^i$ and denote the approximation of the inverse hessian  $\nabla^2 \mathcal J(\x^i) ^{-1}$ as $\Gamma_i$. We note the risk of confusion in the notation between iterative approximations of the entire trajectory and components of the trajectory, all of which have the root notation $\x$. To this end, we denote by $\x$ iterations with entire trajectories, with superscript notation for the iteration count, i.e., $\x^i$, and by $\x_i$ the temporal component $i$ of the trajectory vector $\x$. In the L-BFGS method, the approximation of the inverse Hessian at each iteration is updated by the following equation:

\begin{equation}
\label{eqn:lowrank_update}
    \Gamma_{i} = \left(I - \frac{s_i g_i^T}{g_i^T s_i}\right) \hat{\Gamma}_i \left(I - \frac{g_i s_i^T}{g_i^T s_i}\right) + \frac{s_i s_i^T}{g_i^T s_i} \,.
\end{equation}
where $g_{i}=\nabla \mathcal J(\x^{i}) - \nabla \mathcal J(\x^{i-1})$, $s_{i}=\x^{i}-\x^{i-1}$ and  $\hat{\Gamma}_i$ is a recursively defined previous approximation based on the previous $m$ pairs of update vectors $\{g_j,s_j\}_{j=i-m+1}^i$, $I$ is the identity matrix of appropriate dimensions. Instead of computing and storing $\Gamma_i$, only a history of update vectors is stored, each vector having dimension $d_{x}$. The left product by $\Gamma_i$ can be applied 'matrix-free' using the history of update vectors in a two-loop recursion algorithm \cite[Algorithm 7.4]{nocedal1999numerical}. Thus, using memory $\mathcal O(md_x)$, $x$ is updated with following recursion
$$\x^{i+1} = \x^{i} -\gamma_{i}\Gamma_{i}\nabla \mathcal J(\x^i),$$
where the coefficient $\gamma_i$ is chosen to satisfy Wolfe conditions \cite{nocedal1999numerical}. The approximate minimizer of the cost function (equation \ref{eqn:wc4dvar-cost}), denoted $\x_{\textrm{L-BFGS}}^*$ is defined as the result of L-BFGS after $n_{\textrm{iter}}$ iterations i.e.
$\x_{\textrm{L-BFGS}}^* := \x^{n_{\textrm{iter}}}$. The choice of hyperparameters $n_{\textrm{iter}}$ and $m$ are discussed in \S \ref{ss:hyperparams}.

 To compute the gradients, we utilize the automatic differentiation of FourCastNet. We construct the loss function \ref{eqn:wc4dvar-cost} in pytorch, which includes forward passes through FourCastNet, also implemented in pytorch. This allows us to efficiently compute the gradients in equations  \eqref{eqn:AugLagGrad} using backward propagation. We do not need to explicitly construct and store the adjoints $\mathbf M_{\rm k, k+1}^T$.

The L-BFGS optimization requires an initial guess $\x^0$. To provide it, we take a cubic spline interpolation of each observation $\y_k$ evaluated on the fine grid, which is denoted $\tilde{\y}_k$, and initialize our optimization with $\x^0 = (\x_0^b, \tilde{\y}_1, ..., \tilde{\y}_N)$.



\subsection{Uncertainty Quantification}
\label{sec:UQ}
Since the observation operator is inherently lossy and since our observations are subject to random noise, we do not expect to be able to reconstruct the actual trajectory perfectly. We want to quantify our uncertainty in the reconstructed trajectory given what we do know -the probability distributions of measurement noise and background noise- and what we can estimate - the model error distribution. We do this by approximating the posterior distribution $p\left(\x|\y,\x_0^{\rm b}\right)$, where $\x$ and $\y$ represent the entire trajectory $(\x_0,...\x_N)$ and entire observation sequence $(\y_1,...,\y_N)$ respectively.

\begin{figure}
    \centering
    \includegraphics[width=0.5\linewidth]{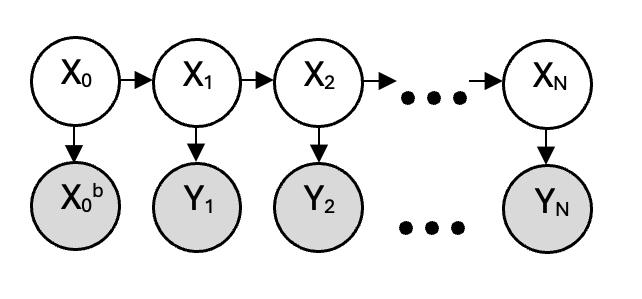}
    \caption{Graphical model denoting conditional dependence of random variables used for estimating the posterior in section \ref{sec:UQ}. Shaded circles represent observed variables, and unshaded circles represent unobserved variables whose posterior we seek to estimate}
    \label{fig:graphical_model}
\end{figure}

If we assume the following probabilistic relationships 
\begin{subequations}
	\label{eqn:cond_dists}
	\begin{eqnarray}
\x_0 &\sim&  \mathcal N(\x_0^{\rm b} ,\mathbf{B_0}),\\
\y_{\rm k}   | \x_{\rm k} &\sim&  \mathcal N\left(\mathcal{H} \left(\x_{\rm k}\right) , \mathbf{R}\right) \quad \textrm{for k} = 1,...,N, \\
\x_{\rm k+1}  | \x_{\rm k} &\sim&  \mathcal N\left(\mathcal{M} \left(\x_{\rm k}\right) , \mathbf{Q}\right) \quad \textrm{for k} = 0,...,N-1
\end{eqnarray}
\end{subequations}
and the conditional independencies of the graphical model in Figure \ref{fig:graphical_model}, then the probability density of the posterior distribution has the following proportionality relationship,
\begin{equation}
\label{eqn:posterior_prop}
    p\left(\x|\y,\x_0^{\rm b}\right) \propto \exp\left(-\J(\x)\right)
\end{equation}
where $\mathcal J$ is the WC4DVAR cost function given by Equation \ref{eqn:wc4dvar-cost}. Note that $\x^*$, a minimizer of $\mathcal J$, is a maximum a posterior (MAP) estimator of $\x$ given observations $\y$ and $\x_0^{\rm b}$ . 

Using a second-order Taylor expansion approximation of $\mathcal{J}$ about $\x^*$, and exploiting that $\nabla \mathcal{J}(\x^*)=0$ due to the optimality conditions, we obtain the following:
\begin{align}
\label{eqn:expansion}
    p\left(\x|\y,\x_0^{\rm b}\right) &\approx k\exp\left(-\mathcal{J}(\x^*)-\nabla \mathcal{J}(\x^*)(\x-\x^*)-\frac12(\x-\x^*)^T\nabla^2\mathcal{J}(\x^*)(\x-\x^*)\right) \\
    \nonumber
    &\approx k'\exp\left(-\frac12(\x-\x^*)^T\nabla^2\mathcal{J}(\x^*)(\x-\x^*)\right).
\end{align}
This gives a Gaussian approximation for the posterior distribution of the trajectory,
\begin{equation}
\label{eqn:posterior_sim}
    \x|\y,\x_0^{\rm b} \sim \mathcal N(\x^*,\nabla ^2\mathcal{J}(\x^*)^{-1}).
\end{equation}

\subsubsection{Sampling Scalably from the Posterior \eqref{eqn:posterior_sim}}
To sample from the posterior distribution (\ref{eqn:posterior_sim}), we utilize the approximate inverse Hessian that is already computed in the optimization step  
$\Gamma \approx \nabla ^2\mathcal{J}(\x^*)^{-1}$;  compare to \eqref{eqn:posterior_sim}.
Samples from the resulting Gaussian distribution can then be drawn by transforming a standard normal distribution $Z\sim\mathcal N(0,I_{d_x(N+1)})$, 
\begin{equation}
    \tilde{\x} = \x^* + \Gamma^{1/2}\mathbf Z.
\end{equation}

Recall that, as used in L-BFGS, $\Gamma$ is stored as a history of update vectors. The matrix-vector product of $\Gamma$ and a vector $\mathbf Z$ can be computed efficiently and matrix-free. To compute efficiently and scalably $\Gamma^{1/2}\mathbf Z$, we use a polynomial approximation of the matrix-square root that can be applied matrix-free to $\mathbf Z$ as follows

\begin{equation}
\label{eqn:polynomial-approx}
    \Gamma^{1/2}\mathbf Z = \sum_{j=1}^{n_p}c_j P_j(\Gamma)\mathbf Z. 
\end{equation}
where $n_p$ is the number of polynomials used in the approximation. Polynomials $P_j$ (of degree $j$) and normalizing coefficients $c_j$ are derived via an orthogonalization procedure following Algorithm 1 in \cite{5a83b8c7658e42ce881add5648f3c7ab}.

\subsection{Ensemble Kalman Filter} \label{ss:ENKF}
While our primary proposed method is to use L-BFGS on the 4DVar functional \eqref{eqn:wc4dvar-cost}, we find it helpful to compare our results with the EnKF method, which in principle can also be used to carry out trajectory estimation with uncertainty. As we developed our approach, it became apparent that to make a fair comparison, we needed to put some effort into circumventing some issues in implementing EnKF at scale, as we did not have access to suitable open-source robust implementations. In this subsection, we give some details of EnKF and some of our algorithmic choices. 
\subsubsection{Method}
The Ensemble Kalman Filter is a data assimilation method that estimates the latent state by sequentially forecasting an ensemble of estimates and updating them with observational data at each time step. Together with Gaussian assumptions found in \ref{eqn:cond_dists}, EnKF is based on the assumption that the conditional distribution $p(\x_i|\y_1,...,\y_{i-1})$ also follows a normal distribution denoted $\mathcal N({\tilde {\mu}_i , \mathbf{C}_i})$. This is known as the background distribution at time $t_i$, and it is estimated from forecasted ensemble members \cite{Katzfuss01102016}. For each time $i = 0,...,N-1$,  EnKF consists of a forecast step and an analysis step to derive the posterior distribution $p(\x_i|\y_1,..., \y_{i-1}\y_{i})$ and corresponding ensemble members. We denote the ensemble of $\nens$ latent-state estimates at time $t_{i}$ by $ 
\{ \x_{i}^{ \langle\ell\rangle}\}_{\ell = 1,\dots,\nens}.
$. These represent samples for the posterior distribution $\x_i|\y_1,...,\y_{i-1}, \y_{i}$ and are recursively defined from the previous time steps as follows:

\textbf{Forecast step}\\
We apply a stochastic forecasting step to produce the background ensemble members at time $t_{i-1}$ denoted $\x_{i-1}^{ {\rm b}\langle\ell\rangle}$
\begin{align}
	\label{eqn:forecast-step}
	& \x_i^{ {\rm b}\langle\ell\rangle} = \mathcal{M}(\x^\ell_{i-1}) + \xi^\ell_{i}, &\xi^\ell_{i} \sim \mathcal N(0,\mathbf Q) 
\end{align}
for $\ell = 1, ..., \nens$ and $\xi^\ell_{i}$ all i.i.d. 

\textbf{Analysis step}\\
The covariance $\mathbf C_i$ is computed from $\{ \x_{i}^{{\rm b}\langle\ell\rangle}\}_{\ell = 1,\dots,\nens}$. How this matrix is computed is discussed in \S \ref{sss:mat_free} and \S \ref{sss:complexity}. The Kalman gain matrix $\mathbf{K}_i $ is defined as follows, 

\begin{equation}
	\label{eqn:kalman-gain}
	\mathbf{K}_i = \mathbf{C}_i\, \HH^\mathrm{T} \left( \HH\, \mathbf{C}_i \, \HH^\mathrm{T} + \R \right)^{-1},
\end{equation}
where $\HH$ is the Jacobian of the observation operator. 

Observations are assimilated to update our estimate at time $t_i$ in the analysis step (Equation \ref{eqn:analysis-step}).
\begin{align}
	\label{eqn:analysis-step}
	&\x^\ell_{i} = \x_i^{ {\rm b}\langle\ell\rangle}+  \mathbf K_i(\y_{i} + \epsilon_{i}^\ell - \mathcal H(\x_i^{ {\rm b}\langle\ell\rangle})), 
	&\epsilon_{i}^\ell \sim \mathcal N(0,\mathbf R)  
\end{align}
for $\ell = 1, ..., \nens$ and $\epsilon^\ell_{i}$ all i.i.d.

%
%
%

\textbf{Initialization}\\
\noindent The $\nens$ members at are initialized at time $t_0$ by 
\begin{align}
	\label{eqn:init_enkf}
	& \x_0^{ \ell} =\x_0^{ b} + \eta^\ell, &\eta^\ell \sim \mathcal N(0,\mathbf B_0) 
\end{align}
for $\ell = 1, ..., \nens$ and $\eta^\ell$ all i.i.d. 

One of the advantages of an EnKF is its built-in approach to uncertainty quantification. The ensemble members are used to estimate the conditional distribution of the fine-scale information conditioned on the coarse scale. Under the linear and Gaussian error assumptions $\x_i^\ell$ are distributed according to the posterior distribution $p(\x_i|x_0^b, \y_1, ... ,\y_i)$. 



%

\subsubsection{Matrix free approach}
\label{sss:mat_free}
In the applications we target here, 
the matrix $\mathbf C_i$ cannot be assembled and stored. This is because it is of dimension $d_x \times d_x$ where $d_x$ is the dimension of the entire fine-scale trajectory. 
The sample background covariance $\hat{\mathbf C}_i$ at time $t_i$ can be approximated from the sample by 
\begin{equation}
\label{eqn:cov-nonlocal}
    \hat{\mathbf C}_i = \frac{1}{\nens-1} \, \sum_{\ell=1}^{\nens}(\x_i^{ {\rm b}\langle\ell\rangle}-\overline{\x}^{\rm b}_i)(\x_i^{ {\rm b}\langle\ell\rangle}-\overline{\x}^{\rm b}_i)^T
\end{equation}

where
\begin{equation}
\overline{\x}^{\rm b}_i = \frac{1}{\nens} \sum_{\ell=1}^{\nens} \x_i^{ {\rm b}\langle\ell\rangle}.
\end{equation}

This  $d_x \times d_x$ matrix is infeasible to assemble or store since in our setting $d_x > 2\times 10^7$. If ${\mathbf C}_i$ is defined according to the sample covariance $\hat{\mathbf C}_i$ in \ref{eqn:cov-nonlocal}, we use the Woodbury matrix identity be used to compute the product by $\mathbf K_i$ with an inverting matrix of only size $\nens \times \nens$. 

\subsubsection{Localization}
Another issue arising from our data's high dimension is the problem of undersampling. A small number of ensemble members relative to the dimension of the state variable, while required by considerations of computational effort and memory, can lead to a poor estimate of the covariance matrix $\mathbf C_i$, which, without additional assumptions, would need from \eqref{eqn:cov-nonlocal},  a number of members comparable to its dimension. For example, the dimensionality of the input and output of FourCastNet is over twenty million, which would require tens of millions of vectors with tens of millions of components, which would be computationally infeasible.  To this end, as it is common in the usage of EnKF in assimilation literature
\cite{DistanceDependentFilteringofBackgroundErrorCovarianceEstimatesinanEnsembleKalmanFilter}
\cite{ASequentialEnsembleKalmanFilterforAtmosphericDataAssimilation},
  we assume that the covariance matrix can be approximated well by one that has local support - i.e. that the correlation between sufficiently distant pairs of points tapers to zero as the distance between them increases. When the true background 
covariance matrix $\mathbf C_i$ has such a property, so a smaller ensemble should estimate local correlations well. 

We then use the following estimated background covariance matrix
\begin{equation}
\label{eqn:localcov}
    \mathbf{C}_i :=  \rho \circ( \mathbf{X}_i\, \mathbf{X}_i^T).
\end{equation}
Here $\mathbf X_i \in \mathbb R^{d_x \times \nens}$ is defined as
\begin{equation}
    \mathbf X_i = \frac{1}{\sqrt{\nens-1}}(\x_i^{ {\rm b}\langle1\rangle} - \overline{\x}^{\rm b}_i, ..., \x_i^{ {\rm b}\langle\ell\rangle} - \overline{\x}^{\rm b}_i,..., \x_i^{ {\rm b}\langle\nens\rangle} - \overline{\x}^{\rm b}_i).
\end{equation}
and $\rho \in \mathbb R^{d_x \times d_x}$ is a 'localization matrix'. $\circ$ represents the Schur product, which is also known as Hadamard or element-wise.

\subsubsection{Localization Structure}
To complete the definition of our scalable covariance estimation, \eqref{eqn:localcov}, we need to specify the choice of localization procedure, also known as tapering in Gaussian process literature. 
In our work, for tapering the sample covariance, we use the Gaspari-Cohn localization function \cite{GaspariCohn}:

\begin{equation}
\label{eqn:rho-definition}
    \rho_{(v_1,p_1,v_2,p_2)} = G\left(\frac{\textrm{hav}(p_1,p_2)}{r_{\rm loc}}\right)\delta_{v_1, v_2}. 
\end{equation}

Here $\rho_{(v_1,p_1,v_2,p_2)}$ is the element of the localization matrix $\rho$ corresponding to the covariance between atmospheric variable $v_1$ at point $p_1$ and $v_2$ at $p_2$; $\delta_{v_1, v_2}$ is the Kronecker delta; $\textrm{hav}(\cdot)$ is the haversine 
distance\cite{sinnott1984virtues}  between two points on the globe;  and $r_{\rm loc}$, the localization radius, is a scaling factor in the tapering. The Gaspari-Cohn tapering function $G$ is defined piecewise as follows.

\begin{equation}
G(r) =
\begin{cases}
    1 - \frac{5}{3}r^2 + \frac{5}{8}r^3 + \frac{1}{2}r^4 - \frac{1}{4}r^5, & 0 \leq r \leq 1, \\
    -\frac{2}{3r} + 4 - 5r+\frac{5}{3}r^2 + \frac{5}{8}r^3 - \frac{1}{2}r^4 + \frac{1}{12}r^5, & 1 < r \leq 2, \\
    0, & r > 2,
\end{cases}
\end{equation}
The resulting tapering matrix  $\rho$ is a symmetric positive definite matrix, and thus $\mathbf C_i$ is symmetric and positive definite by the Schur product theorem \cite{STYAN1973217}.

\subsubsection{Memory and Computational Considerations}
\label{sss:complexity}
While using this localization will help mitigate the undersampling problem, the tapered covariance matrix is no longer low rank, and the Woodbury matrix identity can no longer be used to compute the analysis step 'matrix-free.' Nevertheless, certain assumptions about the localization, measurement error, and observation operator allow us to significantly reduce the computational and memory requirements of the analysis step. We present how we exploit these features to accelerate the computations. 

\textbf{Assumption 1.}
The matrix $\mathbf R$ is sparse and $\mathbf R_{ij} = 0$  if $\rho_{ij} = 0$.

\textbf{Assumption 2.}
 We assume that the observation operator simply returns the values on a coarser grid than that of $\x$ with no smoothing, i.e that $\mathcal H(x) = \HH x$ and the rows of $\HH$ are distinct Cartesian basis vectors.

 Though matrix multiplication does not, in general, distribute over the Schur product, observe that 
\begin{equation}
(A\circ B)\mathbf e_j = (A\mathbf e_j) \circ (B\mathbf e_j) \quad \textrm{and}\quad \mathbf e_i^T (A\circ B)= (\mathbf e_i^T A) \circ (\mathbf e_i^T B)   
\end{equation}
for any matrices $A$ and $B$ and Cartesian basis vectors $\mathbf e_j$, $\mathbf e_i$ with dimensions, the products are all well defined. Thus 
\begin{equation}
    (\rho \circ( \mathbf{X}_i\, \mathbf{X}_i^T))\HH^T = \mathcal H(\rho)^T\circ(\mathbf X_i \mathcal H(\mathbf X_i)^T)
    \label{eqn:sparse-product}
\end{equation}
and 
\begin{equation}
    \HH(\rho \circ( \mathbf{X}_i\, \mathbf{X}_i^T))\HH^T = \mathcal H(\mathcal H(\rho)^T)\circ(\mathcal H(\mathbf X_i) \mathcal H(\mathbf X_i)^T).
\end{equation}

In equation \ref{eqn:sparse-product}, the product of $\mathbf X_i$ and  $\mathcal H(\mathbf X_i)^T$ is computed according to the sparsity pattern of $\mathcal H(\rho)^T$. 

Due to the assumed structure of $\rho$, different atmospheric variables have covariance 0. This allows us to separate $\mathbf C_i$ into blocks by different atmospheric variables, e.g., pressure and humidity. Furthermore, 
the tapering ensures that all variables at points at a distance greater than $2r_{\rm loc}$ also have covariance 0.
Thus, we can divide the latitudes into b zonal bands such that points in each band correlate only with points in the adjacent zonal bands.
This allows us to formulate $\HH \mathbf C_i\HH^T + \mathbf R$ as a block tridiagonal matrix with $n_v b$ blocks on the diagonal each of size $\frac{d_y}{n_vb}$. 


We use the Thomas algorithm\cite{ContedeBoor} 
for Gaussian elimination of tridiagonal systems to solve
\begin{equation}
\label{eqn:enkf-inv-step}
    (\HH \mathbf C_i\HH^T + \mathbf R)\mathbf v^\ell = \y_{i} + \epsilon_{i}^\ell - \mathcal H(\x_i^{ {\rm b}\langle\ell\rangle})
\end{equation}
for $\mathbf v^\ell$. 
The Thomas algorithm has linear complexity\cite{ContedeBoor}.  
When applied to block-tridiagonal matrices, it has linear computational complexity in the number of blocks and cubic complexity in block size. Thus the computational complexity is $$\mathcal O\left(n_vb\left(\frac{d_y}{n_vb}\right)^3\right) \sim \mathcal O\left(\frac{d_y^3}{(n_vb)^2}\right)$$ greatly reduced compared to the $\mathcal O(d_y^3)$ complexity of inverting arbitrary $d_y\times d_y$ matrix. 

Due to the block diagonal structure of the background covariance  $\mathbf C_i$ relative to the atmospheric variables, we can solve the systems corresponding to each one separately. The Thomas algorithm for each atmospheric variable has the memory complexity $\mathcal{O}(\frac{d_y^2}{bn_v^2})$

$$\mathbf K(\y_{i} + \epsilon_{i}^\ell - \mathcal H(\x_i^{ {\rm b}\langle\ell\rangle})) = \mathbf C_i\HH^T\mathbf v^\ell = (\mathcal H(\rho)^T\circ(\mathbf X_i \mathcal H(\mathbf X_i)^T))\mathbf v^\ell$$
where $\mathbf v^\ell$ is the solution of Equation \ref{eqn:enkf-inv-step}.

Each row of $(\mathbf X_i \mathcal H(\mathbf X_i)^T)$ can be computed according to the sparsity pattern of the row of $H(\rho)^T$ before taking the product with $\mathbf v$. Assuming that $\mathcal H$ can be applied matrix-free, without needing to construct the $\HH$ matrix, this product in equation \ref{eqn:sparse-product} requires the memory linear in the number of non-zero values in $\mathcal H(\rho)$ divided by the number of atmospheric variables since $\mathcal H(\rho)$ consists of $n_v$ identical blocks.

We have that the memory usage is $\sim \mathcal O(\textrm{nnz}(H(\rho))/n_v)$ $\sim \mathcal O(\frac{d_xd_y}{b^2n_v^2})$,
where nnz denotes the number of non-zeros in a sparse matrix. Since $b$ is inversely proportional to localization radius $r_{\rm loc}$, the localization radius must be carefully chosen to maximize the solution's accuracy and balance memory limitations.


\section{Numerical Experiments}
\label{sec:experiments}
\begin{figure}[ht]
	\centering
	\includegraphics[width=0.4\linewidth]{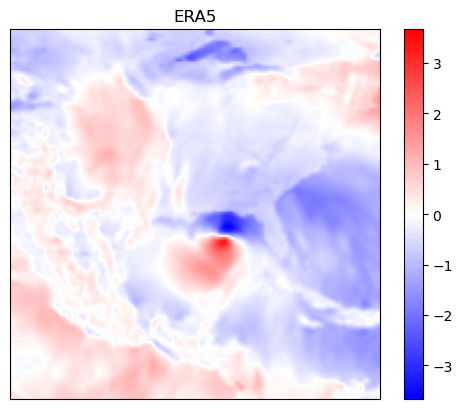}
	\includegraphics[width=0.4\linewidth]{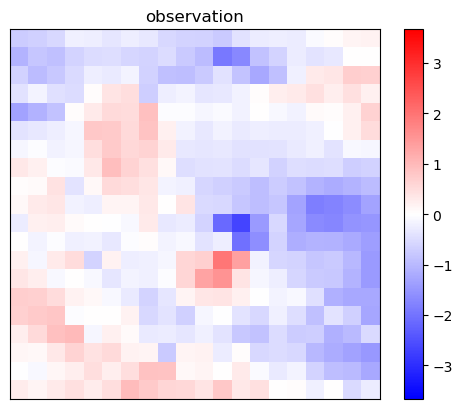}
	\caption{Normalized $U_{10}$ zonal wind variable during Hurricane Michael at 06:00 October 9th UTC on a subregion from 70$^\circ$W to 110$^\circ$W and 10$^\circ$N to 50$^\circ$N The left is the ground-truth ERA5 fine-scale $\x_k$. The right is the coarse and noisy observation $\y_k$}. 
	\label{fig:fine_and_coarse}
\end{figure}

We apply our methodology to downscale a coarse and noisy observation of 2018 Hurricane Michael. We consider the 78-hour time window from 00:00UTC October 7th  to 06:00UTC October 10th, 2018, corresponding approximately to the three days leading up to the hurricane's landfall. We use the ERA5 reanalysis dataset\cite{ERA5}, which we treat as ground truth for our numerical experiments. This is used to generate the coarse observed state $\y$ and to validate the results. A visualization of the ground-truth fine scale and corresponding observation is displayed in figure \ref{fig:fine_and_coarse}. We use a pretrained FourCastNet\cite{FourCastNetGithub}, which was trained from 1979 to 2015, so Hurricane Michael is not included in the training set. We also use ERA5 years 2016 and 2017 for hyperparameter tuning and covariance estimation.

Let $\x^{\rm{true}}$ denote the ground-truth fine scale state taken from ERA5 that we attempt to recover. $\x^{\rm{true}}$ is composed of the 20 atmospheric variables described in table \ref{tab:vars} at a 6 hour intervals and at $0.25^{\circ}\times0.25^{\circ}$ spatial resolution. The time window we consider corresponds to 14 temporal points. Thus, the dimension of state variable $\x$ representing the entire fine scale trajectory is $$\textrm{dim}(\x) = 14d_x = 290,304,000.$$

The observed state $\y$ is generated by taking a coarse and noisy observation of $\x^{\rm{true}}$. Specifically,  $\y$ consists of all 20 variables at each time step on a $2^{\circ}\times2^{\circ}$ (90$\times$180) subgrid with the addition Gaussian noise. This corresponds to every eighth zonal grid line and every eighth meridional grid line in the fine grid. The coarse observation has dimensions $d_y :=  \textrm{dim}(\y_i) = 20 \times 90 \times 180 = 324000$; that is $64$ times fewer spatial points than what we are trying to reconstruct:

\begin{equation}
\label{eqn:yobs_def}
\y_k  := \mathcal H(\x_k^{\rm{true}})  + \mathbf{\nu}_k,
\end{equation} 
where  $\mathbf{\nu}_k$ are drawn from $\mathcal N\left(\mathbf 0 ,0.01^2 I_{d_y}\right)$  i.i.d. for k = 1,...,N. The choice of this Gaussian noise is discussed further in  \S  \ref{sss:measerr}.

The observation operator, which returns the subgrid values is defined as $\mathcal{H} : \mathbf{v} \mapsto \mathbf{H}\mathbf{v}$, where  $\mathbf{H}\in \mathbb R^{d_y\times d_x}$ is a matrix whose rows are $d_y$ distinct standard basis vectors of $\mathbb R^{d_x}$ corresponding to the points on the subgrid. This implies assumption 2 in \S \ref{sss:complexity}.

\subsection{Error Covariance Matrices}
\label{sec:error-covs}
For both WC4DVAR and EnKF, several error covariances must be prespecified, namely model error covariance $\mathbf Q_k$, measurement error covariances $\mathbf R_k$, and background error covariance $\mathbf B_0$. The mean of the background distribution $\x_0^b$ must also be supplied. We describe here how each of these is chosen. We use the same covariances and background mean choices for WC4DVAR and the EnKF methods. 

\subsubsection{Model error covariance} \label{sss:moderr}
As mentioned, we assume that the model error covariance matrix $\mathbf{Q}$ is time-independent. To estimate $\mathbf{Q}$, we use a subset of ERA5 data from 2016. This disjoints from the FourCastNet training and Hurricane Michael test sets. We used 100 times drawn uniformly at random from ERA5 year 2016. The difference between the single timestep (6h) FourCastNet forecast of each of these and the ERA5 states 6 hours after was used to estimate the model error variances according to the method described in \ref{sec:Q_est}. The empirically computed standard deviations for model error in each normalized atmospheric variable take values between 0.01 and 0.25. 

\subsubsection{Measurement error covariance}
 \label{sss:measerr}
The choice of matrices $\mathbf R_{\rm{k}}$ encode our confidence in the observations. In climate model contexts, these choices would be informed by our confidence in the climate model output. These can be computed with a method similar to our method for computing $\mathbf Q$ given a climate model. However, in many practical contexts, GCMs will output an ensemble of scenarios, each of which we want to downscale and which are not a simple Gaussian distribution centered on an actual state. Thus, one would assign high confidence to each GCM output and downscale each output separately. 

To this end, for our numerical experiments, we want to choose measurement errors that are small compared to FourCastNet model errors, also accounting for the fact that the ERA5 data is the ``ground truth,'' which we use when we assimilate. We decided on a small measurement error that is comparable to the smallest model error standard deviation, which is $0.01$, or, equivalently, a variance of 0.0001. Since the ERA5 data is the ``ground truth'', choosing a small variance means that we trust the observations more and this will be reflected in the cost function \eqref{eqn:wc4dvar-cost} -- where the inverse of the variance will weight the second term. 
We choose diagonal and time independent $\mathbf R$ and use this matrix for generating noisy observations as well as for WC4DVAR (Equation \ref{eqn:wc4dvar-cost}) and EnKF (Equation \ref{eqn:analysis-step}). For the reasons above, we chose the measurement error variance matrix as 
$$\mathbf R = 0.0001I_{d_y}$$ so that observations are considered to have errors comparable to the most accurate components of the FourCastNet model \S \ref{sss:moderr}. 
We use the same scale of measurement error in our assimilation methods as we do to produce $\y_k$. In operational settings, when one might want to tune the $\R$ matrix, one can use Desroziers' method for assessing error misspecification. The DA community uses a similar approach in scenarios where $\mathbf R$ is unknown and to generate synthetic observations \cite{Cioaca_2012,Cioaca_2014}.

\subsubsection{Background distribution}
Both WC4DVar and EnKF require an estimate of the fine-scale state at the initial time $\x_0^b$. Since we don't directly observe fine-scale information, we want to choose a large background error covariance matrix to assign a low weight to this initial guess.

We choose $\x_0^b = \mathcal S(y_0)$ for a spherical cubic spline interpolation operator $\mathcal S$. In practice, this is close to an unbiased estimator of  $\x^\textrm{true}_0$.  We compute the variance of this estimator by computing background errors on historical data.
$$\eta_t = x_t^{\textrm{true}}-\mathcal S(\mathcal H(x_t^{\textrm{true}}) + \epsilon_t) \quad \textrm{for i.i.d} \quad \epsilon_t \sim \mathcal N(\mathbf{0}, \R)$$
We find the variance is 0.024. We choose $$B = 0.04I_{d_x}$$ to give greater weight to the FourCastNet model than the prior estimate. 











Considering larger measurement and background variances and off-diagonal elements can be beneficial, especially in extreme events where coarse observations attenuate extrema. Additionally, the smoothing we apply to the initial observation will introduce spatially correlated background errors, so a non-diagonal $\B$ matrix could be more informative.

\subsection{Hyperparameters}
\label{ss:hyperparams}
In the L-BFGS optimization of the WC4DVAR cost function (see \S \ref{ss:optimization}), the number of iterations $n_{\textrm{iter}}$  was chosen to be 25, and the history size $m$ was chosen to be 10. The cost function did not stabilize after 25 steps, nor did the actual mean-squared error. We expect a higher accuracy to be achieved with greater history size and number of iterations; however, we chose these values as a good compromise between computational expenditure and overall accuracy.  
The number of polynomials $n_p$ used in the square root approximation (equation \ref{eqn:polynomial-approx}) was 5. 
We sampled ten trajectories from the posterior estimation. 

For the EnKF, the number of ensemble members $\nens$ was set to 100. The ensemble was initialized with a 100 members drawn i.i.d from $\mathcal N(\x_0^{\rm true},\mathbf{B}_0)$.
The estimated background covariance $\mathbf C_i$ in the analysis step (equation \ref{eqn:kalman-gain}) is given by the localized sample covariance with the Gaspari-Cohn localization (equation \ref{eqn:localcov}). The localization radius $r_{\rm loc}$ in equation \ref{eqn:rho-definition} was chosen to be 666.5km, i.e., covariance tapers to zero at 1333km. Entries of $\rho$ corresponding to covariances between different atmospheric variables were set to zero. We did not apply any covariance inflation.



\subsection{Accuracy Comparison}
\label{sec:accuracy_comparison}
The Root Mean Squared Error (RMSE) is used to measure the accuracy of our estimated fine-scale trajectory. Each time step, this is computed on the normalized data across all features and grid points. For a trajectory $\x^* = (\x_0^* ,..., \x_N^*)$, the RMSE at time step $k$ is defined as,
\begin{equation}
\label{eqn:rmse}
    \textrm{RMSE}_k = \frac1{\sqrt{d_x}}\left\|\x_k^* - \x_k^{true} \right\|_2
\end{equation}
where $\x_k^{true}$ is the ground truth ERA5 data at time $k$. We also compared our predictions to an unassimilated, and therefore, unstabilized, ensemble forecast produced using FourCastNet alone. The unassimilated forecast is computed by repeated applications of FourCastNet with no dependence on the observations $\y_1, ... \y_N$ on 100 initial $\x_0$ values drawn i.i.d. from $ \mathcal N(\x_0^b,\mathbf{B_0})$.

\begin{figure}[ht]
    \centering
    \includegraphics[width=0.8\linewidth]{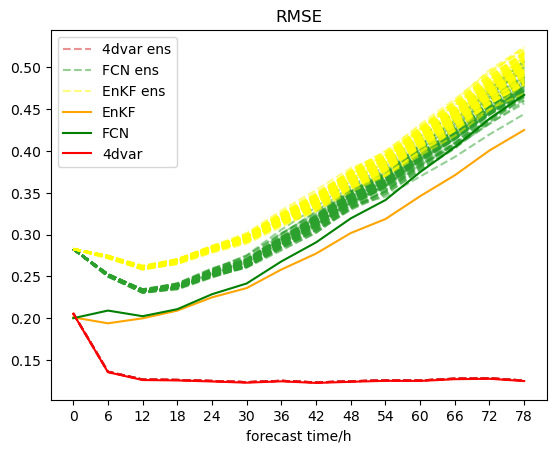}
    \caption{Comparison of RMSE (as defined by \ref{eqn:rmse}) for the WC4DVAR, EnKF, and unassimilated forecasts. }
    \label{fig:RMSE_comp}
\end{figure}

In figure \ref{fig:RMSE_comp}, we compare the RMSE for the estimated fine-scale trajectories using 4DVar, EnKF, and FourCastNet alone, as well as the RMSE for the ensemble members of each method. After the first two steps, the forecast error in the FourCastNet prediction grows with each time step. The ensembles corresponding to perturbed initial values generally show greater RMSE than the forecast initialized on $\x_0^b$. 

The EnKF ensemble mean performs better in RMSE than an unassimilated forecast. The ensemble members generally have a higher RMSE than the unassimilated ensemble forecast, but the ensemble mean is lower than the FourCastNet prediction. RMSE as a metric can reward smoothing, and thus EnKF ensemble mean's lower RMSE may be indicative of smoothed-out fine-scale features and fewer large deviations that may be found in individual ensemble members. 

The WC4DVar stabilizes the forecast. After time 0, the RMSE of WC4DVar shrinks and remains below the $0.2$, the standard deviation of the initial background error, for the remainder of the assimilation period. The ensemble members have very similar RMSE values. This stabilization of RMSE is crucial for downscaling more extended periods. In the operational use of WC4DVAR, the estimated state over a time window can be used as the background estimate for the following window to assimilate longer time horizons. \cite{longwindow4DVAR} We tested our method on a 3-day window. Since our RMSE values are stabilized in time and further fall below the initial background error, the result of the assimilation can be used in a 'sliding window' fashion to downscale longer time windows.

\subsection{Fidelity comparison}
\begin{figure}[ht]
    \centering
    \includegraphics[width=0.23\linewidth]{obs.png}
    \includegraphics[width=0.23\linewidth]{era5.png}
    \includegraphics[width=0.23\linewidth]{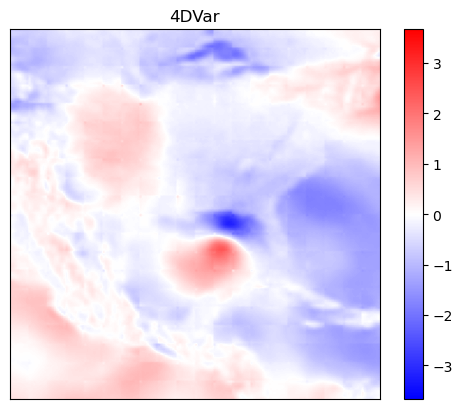}
    \includegraphics[width=0.23\linewidth]{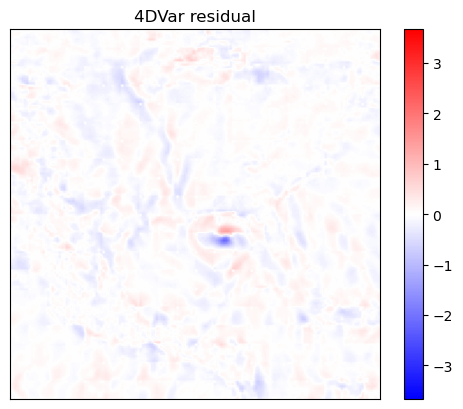}
    \caption{Plot of zonal wind (normalized) at 06:00UTC October 9th, 2018. Left to right are the observation ($\y$), the ground truth  ($\x^{\rm true}$), WC4DVAR result ($\x^*$) and the residual ($\x^*-\x^{\rm true}$)  Each figure uses the same color scale and depicts the same sub-region from 70$^\circ$W to 110$^\circ$W and 10$^\circ$N to 50$^\circ$N.}
    \label{fig:4Dvar_qual_comp}
\end{figure}

\begin{figure}[ht]
    \centering
    \includegraphics[width=0.3\linewidth]{obs.png}
    \includegraphics[width=0.3\linewidth]{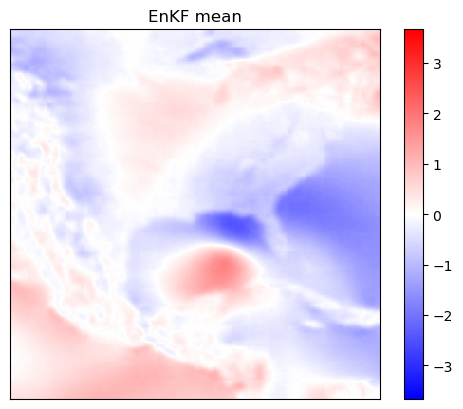}
    \includegraphics[width=0.3\linewidth]{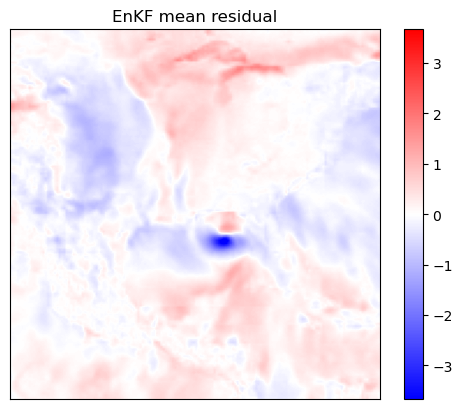}
    \caption{Plot of zonal wind (normalized) at time 06:00UTC October 9th 2018. Left to right are the observation ($\y$), the EnKF result ($\x^{\rm EnKF}$) and the residual ($\x^{\rm EnKF}-\x^{\rm true}$)  Each figure uses the same color scale and depicts the same sub-region from 70$^\circ$W to 110$^\circ$W and 10$^\circ$N to 50$^\circ$N.}
    \label{fig:EnKF_qual_comp}
\end{figure}

We can qualitatively assess the reconstruction of fine-scale details by comparing the results to the ground truth across a spatial subregion, in Figures \ref{fig:4Dvar_qual_comp} and \ref{fig:EnKF_qual_comp}, we plot the normalized U10 variable (zonal wind at 10 meters from the surface) for comparison.

 The WC4DVar recovers much of the fine-scale structures, such as the hurricane's shape and size. The boundaries corresponding to the difference in wind field on land versus sea can be seen in the reconstruction, including for Caribbean islands. When plotted on the same scale, the residual is small for most regions except in the area of most extreme zonal winds, where they are underestimated. 

The EnKF ensemble reconstructs much of the large-scale details but displays significant smoothing of finer details. The hurricane's shape and wind speed show significant errors.

\subsection{Hurricane tracking}

We also track two quantities of interest that are especially relevant when modeling cyclones: maximum wind speed at the surface and minimum mean sea level pressure (MSLP). Since atmospheric pressure is minimized in the eye, the location of minimum MSLP is used to parametrize the track of a cyclone \cite{AnAnalyticModeloftheWindandPressureProfilesinHurricanes}. A bounded region of interest is between $90^\circ$W and $100^\circ$W and between $14^\circ$N and $40^\circ$N, which contains the eye of Hurricane Michael over our assimilation window. We plot the minimum mean sea level pressure attained on the grid for each trajectory. We also plot the maximum over the region of the wind speed at 10m above the surface, computed as $S_{10} = \sqrt{U_{10}^2+V_{10}^2}$ at each point.


Hurricane-tracking is defined by the set of latitude and longitude pairs that minimize the mslp within the region at each timestep, $(i_{\rm k},j_{\rm k}) = \arg\min_{(i,j)} \x_{\rm{k}, mslp, i, j}$.

\begin{figure}[ht]
    \centering
    \includegraphics[width=0.4\linewidth]{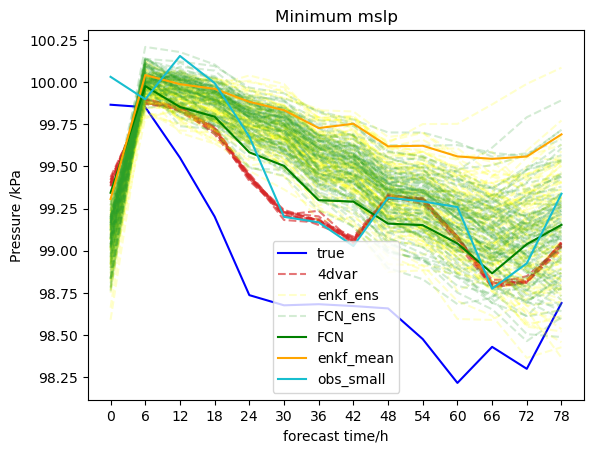}
    \includegraphics[width=0.4\linewidth]{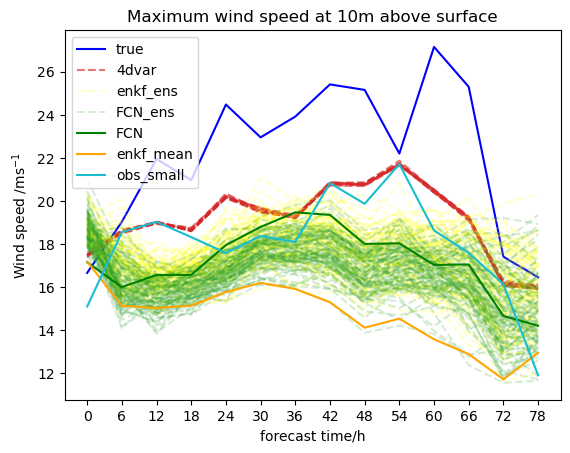}
    \caption{Comparison between different methods of measuring Hurricane Michael extreme values. Left: minimum MSLP. Right: maximum near-surface wind speed. Extrema are taken over the bounded region between $90^\circ$W and $100^\circ$W and between $14^\circ$N and $40^\circ$N. The green and yellow dashed lines represent the ensemble members for the unassimilated and EnKF forecasts, respectively. The red dashed lines represent the samples from the WC4DVAR estimated posterior. The cyan line represents the observation $\y$, and the blue line is the ERA5 ground truth $\x^{\rm{true}}$.}
      \label{fig:minsandmaxes}
\end{figure}

In figure \ref{fig:minsandmaxes}, we compare the models' prediction of these extrema with the true values and the extrema attained on the corresponding subregion of $\y$.  
The maximum wind speed observed on the coarse grid is consistently below the ground truth maximum since the location where the ground truth attains the maximum is unlikely to coincide with the coarse grid. The 4DVar trajectory does improve the estimate of the maximum wind speed, including inferring the existence of a wind speed higher than any observed on the coarse grid. 

The EnKF ensemble members and unassimilated forecast both underestimate this maximum. The EnKF ensemble underestimates the maximum wind more than any ensemble member. Taking the mean of the ensembles smooths out the extrema that might be found in different locations for each ensemble member. For similar reasons, the trajectories overestimate the minimum MSLP. 

\begin{figure}[!ht]
    \centering
    \includegraphics[width=0.5\linewidth]{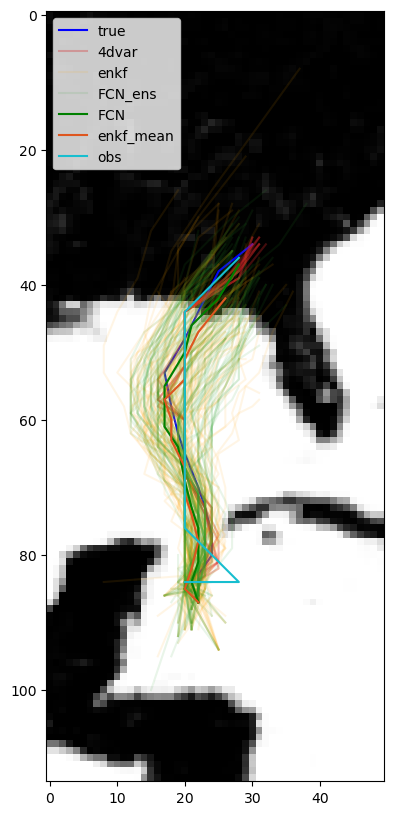}
    \caption{Comparison of Hurricane Michael tracks from the various downscaling methods with the ground truth ERA5 track.}
\label{fig:ens_comp}
\end{figure}

In figure \ref{fig:ens_comp}, we compare the hurricane tracks given by the outputs of each downscaling method. The posterior samples from the WC4DVAR show much greater alignment with the actual track than using the observations on the coarse grid. Additionally, the WC4VAR tracks show greater accuracy and smaller spread than the EnKF ensemble, using FourCastNet alone.

\newpage

\section{Conclusion}

In this work, we presented a novel approach for dynamic downscaling by leveraging data-driven surrogate models, specifically FourCastNet, within a weak-constrained 4DVar framework. While accurate, traditional numerical weather prediction (NWP) models are computationally expensive and time-consuming. Although data-driven models like FourCastNet offer significant computational advantages, they suffer from increasing divergence over longer forecast windows. By integrating FourCastNet into a 4DVar data assimilation framework to carry out downscaling, we stabilized its performance, accounting for model errors and improving forecast accuracy.

Our approach was validated through a hurricane tracking problem using ERA5 data. The results demonstrated superior performance compared to the ensemble Kalman filter (EnKF), both in terms of forecast precision and the ability to quantify uncertainty. Notably, the weak-constraint 4DVar formulation enabled explicit modeling of inherent uncertainties, which is critical for reliable forecasting. This work highlights the potential of combining machine learning surrogates with robust data assimilation techniques to achieve accurate, computationally efficient downscaling.

\subsection{Real-World Applications}

The framework developed in this paper has far-reaching implications for weather forecasting and climate modeling. The enhanced ability to downscale coarse data efficiently and accurately can significantly benefit disaster management, especially in tracking and predicting extreme weather events like hurricanes. Faster and more reliable forecasts can inform emergency response strategies, potentially mitigating damage and saving lives.

Beyond hurricane forecasting, this approach can be extended to other domains requiring high-resolution atmospheric data, such as hydrological modeling, agriculture, and renewable energy forecasting. The twenty prognostic variables, which are tracked at high resolution on the global domain -including temperature, humidity, and total column water vapor- can be used for a myriad of applications, including precipitation prediction.  

The uncertainty quantification capabilities embedded in the framework also make it suitable for risk assessment applications, where understanding the statistics and confidence levels is crucial. The improved computational scalability offered by data-driven surrogate models enables more tractable downscaling of ensembles of GCM outputs. When combined with posterior distribution sampling, this approach facilitates weather generation, including conditional weather generation based on future climate scenarios and GCM parameters. Such capabilities are crucial for understanding the distribution of future weather events, particularly extreme weather, and for producing model output statistics, such as return curves on fine-scale or multi-site extreme events.

Reduced computational costs open avenues for deploying such tools in resource-constrained environments and developing regions. Using a pretrained fine-scale global model also increases accessibility to downscaling in areas where RCMs may not be available. 

\subsection{Future Work and Outlook}

While our proposed framework demonstrates promising results, several avenues for future research remain further to enhance its robustness, accuracy, and applicability. 

One important direction for future work is to investigate how FourCastNet performs under future climate scenarios. A critical question is whether FourCastNet can transfer learned dynamics to new climate regimes. FourCastNet has been shown that it does not extrapolate to out-of-distribution 'gray swan' tropical cyclones when intense tropical cyclones are removed from the training set. However, it has some success generalizing across basins and when only Atlantic basin cyclones are removed from the training set.\cite{sun2024aiweathermodelspredict} 


Our methodology is not limited to FourCastNet and can be extended to other auto-differentiable machine-learning surrogates. As new and more advanced surrogates become available, their integration into the weak-constrained 4D-Var framework may yield improvements in accuracy, uncertainty quantification, and computational efficiency.


Another promising direction is to explore surrogate models that operate at varying resolutions across different spatial regions. In certain areas, high-resolution historical reanalysis data may be available, and training surrogates to handle such heterogeneous resolutions could improve local forecast accuracy. Many such datasets exist, such as the 5.5km resolution Copernicus European Regional Reanalysis.\cite{https://doi.org/10.1002/qj.4764} The weak-constrained 4D-Var framework can naturally incorporate these adaptations by modifying the observation operator and associated error structures.



While the optimization strategy employed in our 4D-Var implementation performs well, it is now the most computationally limiting step. We see several ways in which it could be refined. Enhancements in the approximation of the inverse Hessian could lead to significant improvements in accuracy, speed, and memory efficiency. Notable techniques from spatial statistics include localization of covariances and assuming sparse conditional covariances, which can yield sparsity in the inverse Hessian and Hessian matrices. Imposing specific structures on these approximations could serve as effective preconditioning strategies, accelerating convergence and enhancing posterior estimation accuracy. Exploring these optimization techniques could make the proposed framework even more practical for real-world applications, where computational resources and time constraints are often significant concerns.


A possible area of inquiry is to extend this method to downscale longer time windows. As noted in section \S \ref{sec:accuracy_comparison}, the errors in the WC4DVar estimate are stabilized over the 3-day assimilation window. Moreover, they remain below the initial background error throughout the assimilation window after the initial time. From this, we expect to incorporate this method into a 'sliding window' data assimilation method typical in the operational use of 4DVAR. \cite{longwindow4DVAR} This would allow us to downscale arbitrarily long time intervals, assuming the availability of sufficiently accurate coarse-scale data.

In summary, while this work provides a robust starting point for integrating data-driven surrogates like FourCastNet into weak-constrained 4DVar frameworks, addressing the outlined areas will be pivotal for unlocking the full potential of this approach. As machine learning models and data assimilation techniques continue to evolve, their synergy promises transformative advancements in weather forecasting and climate modeling.

\section*{Acknowledgment}
This material is based upon work
supported by the U.S. Department of Energy, Office of Science,
Office of Advanced Scientific Computing Research (ASCR) under
Contract DE-AC02-06CH11347. We thank Daniel Sanz-Alonso for continued discussions on the topic.

\bibliographystyle{plainnat}
\bibliography{reference}

\vspace{0.1cm}
\begin{flushright}
	\scriptsize \framebox{\parbox{2.5in}{Government License: The
			submitted manuscript has been created by UChicago Argonne,
			LLC, Operator of Argonne National Laboratory (``Argonne").
			Argonne, a U.S. Department of Energy Office of Science
			laboratory, is operated under Contract
			No. DE-AC02-06CH11357.  The U.S. Government retains for
			itself, and others acting on its behalf, a paid-up
			nonexclusive, irrevocable worldwide license in said
			article to reproduce, prepare derivative works, distribute
			copies to the public, and perform publicly and display
			publicly, by or on behalf of the Government. The Department of Energy will provide public access to these results of federally sponsored research in accordance with the DOE Public Access Plan. http://energy.gov/downloads/doe-public-access-plan. }}
	\normalsize
\end{flushright}	

\end{document}